\documentstyle[12pt,epsf]{article}
\def\o{\over}

\def\Bbra{\Big\langle}
\def\Bket{\Big\rangle}

\def\jhep#1#2#3{{JHEP} {\bf #1} (#2) #3}

\def\hpt#1{{\tt hep-th/#1}}

\def\ni       {\noindent}

\def\lbb     {\left[ }
\def\rbb      {\right] }
\def\comma      { \, , }
\def\period     { \, . }
\def\semiket#1  { \, #1 \, \rangle \, }
\def\del        {  \partial  }

\def\half       {  {1\over 2}  }
\def\abs#1      {  \, \vert #1 \vert \,   }
\def\Im#1    { \, {\rm Im } \, #1  }
\def\Re#1    { \, {\rm Re}  \, #1  }
\def\binom#1#2 { \vecii{ {}_{#1} }{\raisebox{.5ex}{$ {}^{#2} $}} }
\def\sqbinom#1#2 { \Bigl(\begin{array}{c} {}_{#1} 
                       \\ \raisebox{.5ex}{${}^{#2}$} \end{array}\Bigr)^2  }

\def\bfR     { {\bf R}}
\def\bfZ     { {\bf Z}}

\def\bfS     { {\bf S}}

\def\calC    { {\cal C} }

\def\calP    {{\cal P}}

\def\zbar   {\bar{z}}
\def\xbar     {\bar{x}}
\def\mbar     {\bar{m}}
\def\nbar     {\bar{n}}

\def\Jbar   { \bar{J} }

\def\bfS     { {\bf S}}
\def\gammabar  { \bar{\gamma} }

\def\mbar     { \bar{m} }
\def\Up     {\Upsilon}

\def\vecii#1#2      {  \Bigl(\begin{array}{c}#1\\#2\end{array}\Bigr)  }
\def\veciii#1#2#3   {  \left(\begin{array}{c}#1\\#2\\#3\end{array}\right)  }
\def\matrixii#1#2#3#4            {  \Bigl( \begin{array}{cc}#1&#2\\#3&#4
                                       \end{array} \Bigr) }
\def\matrixiii#1#2#3#4#5#6#7#8#9 {  \left(\begin{array}{ccc}#1&#2&#3\\
                                     #4&#5&#6\\#7&#8&#9\end{array}\right)  }
\def\eqb         {  \begin{eqnarray}  }
\def\eqe           {  \end{eqnarray}  }
\def\nn               {  \nonumber  }
\def\sectionnumbering { \setcounter{equation}{0}
         \renewcommand{\theequation}{\arabic{section}.\arabic{equation}}}

\def\mysection#1 { \addtocounter{section}{1} \setcounter{subsection}{0}
                 \sectionnumbering \par \bigskip \par \bigskip \noindent
  \begin{center} {\sc \arabic{section} \quad  #1 } \end{center} 
   \par \medskip}
   
\def\mysubsection#1 {\addtocounter{subsection}{1}
      \par \bigskip \noindent  {\normalsize\it
      \arabic{section}.\arabic{subsection} \quad #1  }
   \par \medskip }

\def\csectionast#1    { \begin{center}
    {\large\bf #1  }   \end{center} \par \bigskip}
\def\titleandfile#1#2   {  \begin{center}{\large\bf #1}\end{center}
                            \par\begin{flushright} #2 \end{flushright}  }

\renewcommand{\thefootnote}{\fnsymbol{footnote}}

\newcommand{\vev}[1]{\left<{#1}\right>}

\newcommand{\mfrac}[2]{{\displaystyle\frac{#1}{#2}}}

\begin{document}
\thispagestyle{empty}
\setcounter{page}{0}

\baselineskip 5mm
\renewcommand{\thefootnote}{\fnsymbol{footnote}}
\hfill\vbox{
\hbox{YITP-01-47}
\hbox{UTHEP-442}
\hbox{hep-th/0105283}
}


\baselineskip 0.8cm
\vskip 17mm
\begin{center}
{\large\bf Operator product expansion in $SL(2)$ conformal 
  field theory}
\end{center}


\vskip 12mm
\baselineskip 0.6cm
\begin{center}
    Kazuo ~Hosomichi\footnote[2]
    {\tt hosomiti@yukawa.kyoto-u.ac.jp},                        \\
       {\it Yukawa Institute for Theoretical Physics}           \\
       {\it Kyoto University, Kyoto 606-8502, Japan}            \\
       \vskip5mm
       {\sl and}
     \vskip5mm
    Yuji ~Satoh\footnote[3]
    {\tt ysatoh@het.ph.tsukuba.ac.jp}                           \\
       {\it Institute of Physics, University of Tsukuba}        \\
       {\it Tsukuba, Ibaraki 305-8571, Japan}
\end{center}


\vskip 17mm
\baselineskip=3.5ex
\begin{center}{\bf Abstract}\end{center}
\par
\bigskip
In the conformal field theories having affine $SL(2)$ symmetry,  
we study the operator product expansion (OPE) involving primary fields 
in highest weight representations. For this purpose, we analyze 
properties of primary fields with definite $SL(2)$ weights, and
calculate their two- and three-point functions.  
Using these correlators, we show that the correct OPE is obtained
when one of the primary fields belongs to the degenerate highest 
weight representation. We briefly comment on the OPE in the 
$SL(2,R)$ WZNW model.
%

\noindent 
\noindent
%
\vfill
\noindent
May~ 2001

\newpage

\renewcommand{\thefootnote}{\arabic{footnote}}
\setcounter{footnote}{0}
\setcounter{section}{0}
\baselineskip = 0.6cm
\pagestyle{plain}

%

\noindent
{\bf 1. } \quad 
The conformal field theories having affine $SL(2)$ symmetry
 have been an interesting topic in recent string theory, since
 the $SL(2)$ symmetry expresses the isometry of the $AdS_3$ target space
  or its Euclidean counterpart known as simplest examples 
exhibiting the holography. The CFT on the Euclidean $AdS_3$, namely, the 
$H_3^+$ WZNW model is now well controlled 
\cite{Gawedzki,Teschner1,Teschner2} and, starting from this, 
one may extract useful results for other models with the affine  
$SL(2)$ symmetry \cite{Giveon-K,Maldacena-O}. 

In this note, we continue this line of studies. Our point here is to 
focus on the primary fields with definite $SL(2)$ weights. 
These are important in dealing with highest weight representations,
since highest weight conditions are expressed by certain relations
between the $SL(2)$ spin and weight. In particular, we present pieces of
properties of the primary fields mentioned above, and calculate 
their two- and three-point functions. Using these correlators, we 
discuss the operator product expansion including primary fields 
in highest weight representations. When one of the primary fields 
belongs to the degenerate highest weight representation, we show
that the correct OPE is obtained. We briefly
comment on the OPE in the $SL(2,R)$ WZNW model.  Our 
analyses may serve also as preparatory steps for further studies.

\vskip 3ex
\noindent
{\bf 2.} \quad  
The $H_3^+$ WZNW model \cite{Gawedzki,Teschner1,Teschner2} 
has the action
\eqb
  S = \frac{k}{\pi}\int d^2z \left[\partial\phi\bar{\partial}\phi
  + e^{2\phi}\partial\bar{\gamma}\bar{\partial}\gamma\right].
  \nn
\eqe
 The primary fields are organized by introducing ``boundary
 coordinates'' $x$ and $\bar{x}$:
\eqb
 \Phi_j(z,x)=\left(|\gamma-x|^2e^{\phi}+e^{-\phi}\right)^{2j}
 \period  \nn
\eqe
They behave as if they were primary fields of conformal weight $-j$
 on the $x$-plane. The global part of the affine $SL(2)$ symmetry
 of the model acts onto them as conformal transformations on the $x$-plane
\cite{deBoer-ORT,Kutasov-S}.
They also have conformal weight $ h \equiv -j(j+1)/(k-2)$, and    
OPE's with the $\hat{sl}_2$ currents,   
\eqb
  J^a(z)\Phi_j(w,x)
     &\sim&  -\frac{D^a\Phi_j(w,x)}{z-w} \comma \nn
\eqe
$$
  D^- \ =  \   \partial_x  \comma \quad 
  D^3 \ = \ x  \partial_x -  j  \comma \quad 
  D^+ \ = \ x^2\partial_x - 2jx \period
$$
The expressions with $ \Jbar^a(\zbar) $ are similar.  
The two- and the three-point functions of these fields are given by
\cite{Teschner1,Teschner2,Ishibashi-OS,Hosomichi-OS}
\begin{eqnarray}
\lefteqn{\vev{\Phi_{j_1}(z_1,x_1)\Phi_{j_2}(z_2,x_2)}} 
 \nonumber \\
 &=& |z_{12}|^{-4h_1}\left[
    A(j_1)\delta^{2}(x_{12})\delta(j_1+j_2+1)
   +B(j_1)|x_{12}|^{4j_1}\delta(j_1-j_2)
   \right], \nonumber \\
 && A(j)~=~-\frac{\pi^3}{(2j+1)^2},~~
    B(j)~=~b^2\pi^2[k^{-1}\Delta(b^2)]^{2j+1}\Delta[-b^2(2j+1)],
 \nonumber \\
\lefteqn{
\prod_{a=1}^3\Phi_{j_a}(z_a,x_a) =
  D(j_a)\prod_{a<b}|z_{ab}|^{-2h_{ab}}|x_{ab}|^{2j_{ab}}, }
  \nn \\
 && D(j_a)~=~ \frac{b^2\pi}{2}
  \frac{[k^{-1}b^{-2b^2}\Delta(b^2)]^{\Sigma j_a+1}
        \Up[b]\Up[-2j_1b]\Up[-2j_2b]\Up[-2j_3b]}
       {\Up[-(\Sigma j_a+1)b]\Up[-j_{12}b]\Up[-j_{23}b]\Up[-j_{31}b]}, \nn
\end{eqnarray}
 where $\Delta(x)=\Gamma(x)/\Gamma(1-x),
 b^{-2}=k-2$ and
 $z_{ab}\equiv z_a-z_b,~ j_{12}\equiv j_1+j_2-j_3$, etc.
 An entire function $\Upsilon$ was introduced in 
\cite{Dorn-O,Zamolodchikov-Z}
 and is characterized by the spectrum of zeroes,
\eqb
  \Up(x)=0 ~~{\rm at}~~
  x = -mb-nb^{-1},~~
  x = (m+1)b+(n+1)b^{-1}~~(m,n\in {\bf Z}_{\ge 0}) \period \nn
\eqe
 Using the above correlators and the $SL(2)$ symmetry, 
we can write down the OPE formula:
\eqb
 && \Phi_{j_1}(z_1,x_1)\Phi_{j_2}(z_2,x_2)
 \stackrel{z_1\rightarrow z_2}{\sim}
 \int_{\cal C}dj_3 \, |z_{12}|^{-2h_{12}}\int d^2x_3
 \prod_{a<b}^3|x_{ab}|^{2j_{ab}}{ D(j_a) \o A(j_3)} \Phi_{-j_3-1}(z_2,x_3). 
   \label{OPE} 
\eqe
 where the $j_3$-integration should be taken over all the normalizable
 representations on the Euclidean $AdS_3$, i.e.,
$\calC = \calP \equiv -\half + i \bfR$, 
 if the two operators both belong to the normalizable representations.
 For generic $j_1$ and $j_2$ we assume that certain deformations
 of contours should be made so as to go around the sequences of poles
 in the integrand and ensure the analyticity in $j_{1,2}$.
 Those poles in the integrand are given by 
\eqb
  \begin{array}{llcll}
    {\rm (1a)} \ j_{12} = \bfS \comma & {\rm (1b)} \ j_{12} = -1-\bfS \,; & &
   {\rm (2a)} \ N = \bfS \comma & {\rm (2b)} \ N = -1-\bfS \,;  \\
   {\rm (3a)} \ j_{13} = \bfS \comma & {\rm (3b)} \ j_{13} = -1 -\bfS \, ; 
    & & 
   {\rm (4a)} \ j_{23}= \bfS \comma & {\rm (4b)} \ j_{23} = -1 -\bfS \, ;
  \end{array}
   \label{j3pol}
\eqe
where $N = \Sigma_{a=1}^3 j_a + 1$ and 
$\bfS = l + nb^{-2}$ $(l,n \in \bfZ_{\geq0})$. 
These originate from the zeroes of $\Up$ functions as well as
 the exponents of $|x_{ab}|$ via
\eqb
  {\rm Res} \, x^{\epsilon-l-1}\bar{x}^{\epsilon-n-1}|_{\epsilon=0}
 ~=~ \frac{\pi}{l!n!}\partial^l\bar{\partial}^n\delta^2(x).
 \label{res}
\eqe

\vskip 3ex
\noindent
{\bf 3.} \quad  
We would like to note that the above OPE formula 
has a semi-classical ($k\to \infty$) limit 
 which agrees with the supergravity analysis on the Euclidean
 $AdS_3$ background.
 Using $\Upsilon(x)\stackrel{b\to 0}{\to}{\rm const}\cdot\Gamma(x/b)^{-1}$
 we obtain the following semi-classical OPE formula,
\begin{eqnarray}
 \Phi_{j_1}(x_1)\Phi_{j_2}(x_2)
 &=& \frac{1}{2}\int_{\calP}
 dj_3\int d^2x_3 \prod_{a<b}|x_{ab}|^{2j_{ab}}
 A(j_3)^{-1}D_0(j_a) \Phi_{-j_3-1}(x_3),
 \nonumber \\
 D_0(j_a) &=& \frac{\pi}{2} 
 \frac
  {\Gamma(-\Sigma j_a-1)\Gamma(-j_{12})\Gamma(-j_{23})\Gamma(-j_{31})}
  {\Gamma(-2j_1)\Gamma(-2j_2)\Gamma(-2j_3-1)} \comma \nn
\end{eqnarray}
 out of which the semi-classical four-point function can be expressed as
\cite{Teschner1}
\begin{eqnarray}
\lefteqn{ \vev{\prod_{a=1}^4\Phi_{j_a}(x_a)} }  \nn \\
 &=&
 -\frac{1}{\pi^2}
 |x_{12}|^{2(j_1+j_2)+1}
 |x_{13}|^{2(j_1-j_2)-1}
 |x_{23}|^{2(-j_1+j_2+j_3-j_4)}
 |x_{24}|^{2(-j_3+j_4)-1}
 |x_{34}|^{2(j_3+j_4)+1}
 \nonumber \\ 
 && \hskip-6mm
 \times \int_{\calP} 
 dj \, (2j+1)D_0(j_1,j_2,j)D_0(j_3,j_4,j)
 |x|^{-2j-1}|F(j_2-j_1-j,j_3-j_4-j;-2j;x)|^2, \nn
\end{eqnarray}
 where $x\equiv x_{12}x_{34}/x_{13}x_{24}$. This is obtained also 
by using the relations of completeness and orthogonality of $\Phi_j(x)$. 

The behavior of the integrand for large $| j |$ is evaluated from 
the asymptotic behavior of the hypergeometric function \cite{Bateman}.
Then, it turns out that the contour of $j$-integration can be closed
 in the left half-plane so that we may replace the $j$-integral
 with the sum over poles at
\eqb
  j = j_1+j_2-l,~~
      j_3+j_4-l \ ~~(l\in {\bf Z}_{\ge 0}). \label{jdisc}
\eqe
 The spectrum of intermediate states thus obtained agrees 
 with the semi-classical result of Liu \cite{Liu}.
This further supports the prescription of the OPE given in 
(\ref{OPE}). This also shows that there are two different expansions
for the same quantity because of the existence of infinitely many 
primary fields: one is by $j \in \calP$ and the other is by
(\ref{jdisc}). Similar phenomena are observed in Liouville 
theory \cite{Moore-SS} and in an $SL(2)$ model \cite{Kato-S}. 

\vskip 3ex
\noindent
{\bf 4.} \quad  
The primary fields with definite $SL(2)$ weights, i.e., eigenvalues of
the zero-modes of $J^3(z)$ and $\Jbar^3(\zbar)$, are  
obtained by Fourier transforming the primaries
 $\Phi_j(z,x)$:
\eqb
  \Phi^j_{m\bar{m}}=\int d^2x \, 
       x^{j+m}\bar{x}^{j+\bar{m}}\Phi_{-j-1}(z,x) \period
\label{mode}
\eqe
The above Fourier transformations are well defined only for
 $m-\bar{m}\in {\bf Z}$. In fact, for the $H_3^+$ model,
$ m+\mbar \in i \bfR$ and $m-\mbar \in \bfZ$. 
These have the OPE's, e.g., 
\eqb
     J^\pm(z) \Phi^j_{m\mbar}(w) \sim  \frac{\mp j+m}{z-w} 
     \Phi^j_{m\pm 1 \, \mbar}
    \comma  \quad 
    J^3(z) \Phi^j_{m\mbar}(w) \sim   \frac{m}{z-w} \Phi^j_{m\mbar}
  \period \label{JPhijmm}
\eqe
For evaluating (\ref{mode}), we use the Mellin transform of 
$(z+1)^{-2(j+1)}$ with $z= |\gamma -x|^2 e^{2\phi}$
and Barnes' representation of the hypergeometric function. 

Here, we introduce a coordinate system $(\tau,\varphi,r)$ 
via
\eqb
 e^\phi = e^{-\tau}\cosh r,~~~
 \gamma = e^{\theta_L}\tanh r,~~~
 \bar{\gamma} = e^{\theta_R}\tanh r~~~~
 (\theta_{L/R}\equiv \tau \pm i\varphi)
  \comma \nn
\eqe
in which the metric reads
\eqb
 ds^2 = \cosh^2 r \, d\tau^2  + dr^2 + \sinh^2r \, d\varphi^2
  \period \nn
\eqe
Then, the explicit expression of $\Phi^j_{m\bar{m}}$
is given by
\begin{eqnarray}
  \Phi_{m\bar{m}}^j &=&
 \frac{\pi\Gamma(j+1+m)\Gamma(j+1-\bar{m})}{\Gamma(m-\bar{m}+1)\Gamma(2j+2)}
 e^{m\theta_L+\bar{m}\theta_R}\cosh^{-m-\bar{m}}r\sinh^{m-\bar{m}}r
 \label{PhiF} \\ && ~~~~\times
 F(-j-\bar{m},j+1-\bar{m};m-\bar{m}+1;-\sinh^2r)~~~(m-\bar{m}\ge 0),
  \nn 
\end{eqnarray}
and those for $m-\bar{m}\le 0$ can be obtained by 
exchanging $m$ and $\bar{m}$ with $m\theta_L+\bar{m}\theta_R$ fixed. 
(See also \cite{Teschner1}.)
$\Phi^j_{m\mbar}$ satisfy the reflection relation,
\begin{eqnarray}
 \Phi^{-j-1}_{m\bar{m}} &=& \frac{2j+1}{\pi}c_{m\bar{m}}^{j}
                  \Phi^{j}_{m\bar{m}},
 \nonumber \\
 c^j_{m\bar{m}} &=& 
\pi\frac{\Gamma(m -j)\Gamma(-\bar{m}-j)\Delta(2j+1)}
              {\Gamma(m+j+1)\Gamma(-\bar{m}+j+1)}. \nn
\end{eqnarray}

For $\phi \to \infty$, the asymptotic behavior of $\Phi_j(z,x)$ 
(for generic $j$) is given by \cite{Kutasov-S}
\eqb
   \Phi_j \ \sim \ e^{2j\phi} |\gamma -x|^{4j} 
    + \cdots
  + \frac{-1}{2j+1}
    e^{-2(j+1)\phi} \delta^2(\gamma-x) + \cdots
  \period \nn
\eqe 
Plugging this into (\ref{mode}), we obtain 
\eqb
   \Phi^j_{m\mbar} \ \sim \ 
  c^{-j-1}_{m\mbar} \gamma^{m-j-1} \gammabar^{\mbar-j-1} e^{-2(j+1)\phi} 
   + \cdots  + \frac{1}{2j+1}
    \gamma^{m+j} \gammabar^{\mbar+j} e^{2j\phi}  + \cdots
  \period \label{Phiasympt}
\eqe 
For $j \in \calP$, the leading contribution comes from both series. 
However, for highest weight representations with
\eqb 
   m, \mbar \ \in \ j - \bfZ_{\geq 0} \comma \quad \mbox{or} \quad 
   m, \mbar \ \in \ -j + \bfZ_{\geq 0} \comma
  \label{HW}
\eqe
the coefficient $c^{-j-1}_{m\mbar}$ and, hence, the 
first series vanish.  This shows that the asymptotic behavior largely
changes for highest weight representations.  
The precise form of the asymptotic behavior can be read off, e.g.,
 from (\ref{PhiF}) by using the expression of the hypergeometric function
around $r \to \infty$.  

For highest weight representations, 
the hypergeometric function in (\ref{PhiF})
reduces to a Jacobi polynomial.  For example, 
for $ m = j - n, \mbar = j-\nbar$ $(n,\nbar \in \bfZ_{\geq0})$, 
we have
\eqb
  \Phi^j_{m\mbar} & = & 
   \frac{\pi\Gamma(m+j+1)}{\Gamma(2j+2)} 
   e^{m\theta_L + \mbar\theta_R} y^{\half(m-\mbar)} 
    (1+y)^{\half(m+\mbar)} 
   n! P_{n}^{(m-\mbar,m+\mbar)}(1+2y) \label{PhiJacobi}    \\
 &=&
  \frac{\pi\Gamma(m+j+1)}{\Gamma(2j+2)} 
   e^{m\theta_L + \mbar\theta_R} y^{-\half(m-\mbar)} 
    (1+y)^{-\half(m+\mbar)} \frac{d^{j-m}}{dy^{j-m}} 
    \lbb  y^{j-\mbar} (1+y)^{j+\mbar}\rbb
  \comma \nn 
\eqe
where $ y = \sinh^2 r$. One can confirm that these expressions are 
symmetric with respect to $m$ and $\mbar$, and valid for both 
$m-\mbar \in \bfZ_{\geq 0}$ and $m-\mbar \in \bfZ_{< 0}$.

When the coordinate $\tau$ is continued as $it = \tau$, so that 
$(t,\varphi,r)$ parametrize the Lorentzian $AdS_3$ or $SL(2,R)$, 
$\Phi^j_{m\mbar}$ represent the wave functions on $SL(2,R)$. Precisely,
they are the matrix elements of the $SL(2,R)$ representations 
of the principal continuous series and highest(lowest) weight 
discrete series for $j \in \calP;  m, \mbar \in \bfR$;  and 
$ j \leq  -\half; m, \mbar \in j - \bfZ_{\geq 0} $ 
$(m, \mbar \in -j + \bfZ_{\geq 0})$;  respectively.
Note that the vanishing of the first series in (\ref{Phiasympt}) due
to the highest weight conditions assures the correct normalizability
of the wave functions.

For highest weight representations, 
$\Phi^j_{m\mbar}$ can be associated with the power series expansions of
(the analytic part of) $\Phi_j$ around $x=0$ or $x= \infty$. To see this,
let us define 
\eqb
  \Phi_{m\bar{m}}^{-j-1,-}(z) &=&
  \oint_0\mfrac{dx}{2\pi i}\mfrac{d\bar{x}}{2\pi i}
  x^{m-j-1}\bar{x}^{\bar{m}-j-1}\Phi_j(z,x) 
  \comma \nn  \\
 \Phi_{m\bar{m}}^{-j-1,+}(z) &=&
  \oint_0\mfrac{dx}{2\pi i}\mfrac{d\bar{x}}{2\pi i}
  x^{j-m-1}\bar{x}^{j-\bar{m}-1}\Phi_j(z,x^{-1}) 
  \comma \nn
\eqe
where $m,\mbar \in j -\bfZ_{\geq 0}$ for  $\Phi_{m\bar{m}}^{-j-1,-}$
and $m,\mbar \in -j +\bfZ_{\geq 0}$ for  $\Phi_{m\bar{m}}^{-j-1,+}$.
Since $\Phi_j$ is a function of a certain combination of the variables, 
derivatives of $x$ can be converted to those of $y$.
We thus obtain the explicit form of $\Phi_{m\bar{m}}^{j,-}$ 
similar to (\ref{PhiJacobi}):   
\eqb
 \Phi_{m\bar{m}}^{-j-1,-} & = & 
   \frac{1}{(j-m)!(j-\mbar)!} \del^{j-m}_x \del^{j-\mbar}_{\xbar} \Phi_j 
   \Bigm{|}_{x=\xbar=0} \label{Phi-Phi}  \\
   &=& \frac{2j+1}{\pi} 
 {\Gamma(-j-m)\Gamma(-j-\mbar) \o 
         \Gamma^2(-2j)    \Gamma(j+1-m)\Gamma(j+1-\mbar)}
  \Phi^{j}_{m\mbar}
  \period \nn
\eqe  
The expression for $\Phi^{-j-1,+}_{m\mbar}$ is obtained in a parallel way
by making use of  the inversion relation 
$|x|^{4j} \Phi_j(x^{-1}) = \Phi_j(x) 
\vert_{(\tau,\varphi) \to (-\tau,-\varphi)}$. 
The final result is the same as above up to the replacement 
$(m,\mbar) \to -(m,\mbar)$ in the coefficient in front of $\Phi^j_{m\mbar}$. 
Because of the factors of the Gamma functions,
$\Phi^{j,\pm}_{m\mbar}$ have OPE's similar to (\ref{JPhijmm}), but with
$j$ and $-j-1$ exchanged.

\vskip 3ex
\noindent
{\bf 5.} \quad  
In order to discuss the OPE of $\Phi^j_{m\mbar}$,
we need their correlation functions. 
The two- and three-point functions are  obtained 
 by Fourier transforming those of $\Phi_j(z,x)$.

 First, the two-point functions are  given by
\eqb
 && \vev{\Phi_{m_1\bar{m}_1}^{j_1}(z_1)\Phi_{m_2\bar{m}_2}^{j_2}(z_2)}
  \label{Phijmm2pt} \\
 && \qquad  = |z_{12}|^{-4h_1}\delta^2(m_1+m_2)
   \left\{ A(j_1)\delta(j_1+j_2+1) + c^{-j_1-1}_{m_1\mbar_1} B(-j_1-1)
 \delta(j_1-j_2) \right\},\nn 
\eqe
where
\eqb
  \delta^2(m)\equiv
  \int d^2x \, x^{m-1}\bar{x}^{\bar{m}-1}
   = 4\pi^2\delta(m+\bar{m})\delta_{m-\bar{m},0} \period \nn
\eqe
If we concentrate, e.g., on Im $j \geq 0$ and Re $j \leq -1/2$, only the 
second term in (\ref{Phijmm2pt}) remains. For the highest weight 
representations in (\ref{HW}), the remaining expression can be reduced to 
that proportional to $\delta_{j_1,j_2}$.

The three-point functions are given by
\begin{eqnarray}
  \vev{\prod_{a=1}^3\Phi_{m_a\bar{m}_a}^{j_a}(z_a)}
 &=& \delta^2(\Sigma m_a)
     \prod_{a<b}|z_{ab}|^{-2h_{ab}} D(-j_a-1)W(j_a;m_a),
   \label{Phijmm3pt}
\end{eqnarray}
 where $W(j_a;m_a)$ is the following integral representing
 the group structure:
\begin{eqnarray}
  W(j_a;m_a) &\equiv&
  \int d^2x_1d^2x_2 \,
   x_1^{j_1+m_1}\bar{x}_1^{j_1+\bar{m}_1}
   x_2^{j_2+m_2}\bar{x}_2^{j_2+\bar{m}} \nn \\
   && \qquad \times \, 
   |1-x_1|^{-2j_{13}-2}|1-x_2|^{-2j_{23}-2}|x_1-x_2|^{-2j_{12}-2}. \nn
\end{eqnarray}
 Generically, the integral $W$ is expressed in terms of the generalized
 hypergeometric function $_3F_2$ and is therefore very complicated
 to evaluate.
 However, when
\eqb
  j_1+m_1 = j_1+\mbar_1  =  0 
  \comma \nn
\eqe
the integral is simplified to 
\eqb
  W(j_a;m_a) &=&
   (-)^{m_3-\mbar_3} \pi^2 
    {\Delta(-N) \Delta(2j_1+1) \o \Delta(1+j_{12}) \Delta(1+j_{13})}
     \prod_{a=2,3} {\Gamma(1+j_a+m_a) \o \Gamma(-j_a-\mbar_a)} 
  \period \label{Fhw}
\eqe
In a special case with $m_a = \mbar_a$, 
the three-point function (\ref{Phijmm3pt}) with (\ref{Fhw}) 
reduces (up to a phase) to the result in \cite{Becker-B}.

These correlators are obtained also by appropriately adapting the approach 
in \cite{Hosomichi-OS} to the present case. 
We do not go into details, though.

\vskip 3ex
\noindent
{\bf 6. } \quad  
Given the two- and three-point functions of $\Phi^j_{m\mbar}$,
we would like to discuss the OPE. Here, we follow the argument 
in \cite{Teschner1,Teschner2}: we start from the OPE in the $H_3^+$ case 
as in (\ref{OPE}), and deform the integration contour for generic cases so as
to go around the poles in the integrand.

Thus, we begin with the following form of the OPE,
\eqb
  \Phi^{j_1}_{m_1\mbar_1}(z_1) \Phi^{j_2}_{m_2\mbar_2}(z_2)
  & \sim & |z_{12}|^{-2h_{12}} 
   \sum_{m_3, \mbar_3} \half \int_{\calC} dj_3 \ Q(j_a;m_a) \ 
    \Phi^{j_3}_{m_3 \mbar_3}(z_2) \comma \label{PhijmmOPE}
\eqe
with $\calC = \calP$
for $j_{1,2} \in \calP$, $m+\mbar \in i \bfR$, $m-\mbar \in \bfZ$.
$Q$ is obtained from the consistency 
with the two- and three-point functions through
\eqb
   &&  \Bbra \Phi^{j_1}_{m_1\mbar_1}(z_1) \Phi^{j_2}_{m_2\mbar_2}(z_2) 
    \Phi^{j_4}_{m_4\mbar_4}(z_4) \Bket \nn \\
  && \qquad  \sim \ |z_{12}|^{-2h_{12}} 
   \sum_{m_3, \mbar_3} \half \int_{\calC} dj_3 \ Q(j_a;m_a) \ 
    \Bbra \Phi^{j_3}_{m_3 \mbar_3}(z_2) \Phi^{j_4}_{m_4\mbar_4}(z_4)
   \Bket \period \label{Q23}
\eqe
When one of the primary fields is in the highest weight representation
satisfying $m = \mbar = -j$, we can use (\ref{Phijmm2pt})-(\ref{Fhw}). 
Assuming an appropriate deformation of the contour, we find
that 
\eqb
  Q(j_a;m_a ) &=& \delta^2(m_1+m_2-m_3) 
  { \Gamma(j_2+m_2+1) \o \Gamma(-j_2-\mbar_2)} 
  { \Gamma(-j_3-\mbar_3) \o \Gamma(j_3+m_3+1)} \nn \\
   && \qquad \times \ 
  {\Delta (-2j_2) \Delta(j_{23}+1)\o  R(j_1) R(j_2) A(j_3)} D(j_a)
  \comma \label{Qja}
\eqe
with $R(j) = B(j)/A(j)$.
In the above, we have repeatedly used a formula
\eqb
  D(j_1,j_2,j_3) &=& \pi \Delta(-j_{13}) \Delta(-j_{23}) \Delta(2j_3+1) 
   R(j_3) D(j_1,j_2,-j_3-1) \period \nn
\eqe
Note that the two terms proportional to $\delta(j_3-j_4)$ and 
$\delta(j_3+j_4+1)$ in 
$\Bbra \Phi^{j_3}_{m_3 \mbar_3} \Phi^{j_4}_{m_4\mbar_4}\Bket$ give 
the same contributions to (\ref{Q23}), because
\eqb
    &&  Q(j_1,j_2,j_3;m_a) c^{-j_3-1}_{m_3\mbar_3} B(-j_3-1)
   \ = \  A(j_3) Q(j_1,j_2,-j_3-1;m_a) \period \nn
\eqe

There are two types of poles in $ j_3 $ in $ Q(j_a;m_a) $:
one is $ m $-independent and the other is $ m $-dependent. 
The $ m $-independent poles develop at  
$ j_3 \in (-m_3 +\bfZ_{\geq 0}) \cap (-\mbar_3 +\bfZ_{\geq 0})$.
The $ m $-independent poles come from $ D(j_a) $ 
and $\Delta(j_{23}+1)$. Taking into account the 
zeros in the latter,  we find the $ m $-independent poles in $ Q $ at 
\eqb
  \begin{array}{llcll}
    {\rm (1a)} \ j_{12} = \bfS \comma & {\rm (1b)} \ j_{12} = \bfS' \,; & &
   {\rm (2a)} \ N = \bfS \comma & {\rm (2b)} \ N = \bfS' \,;  \\
   {\rm (3a)} \ j_{13} = \bfS \comma & {\rm (3b)} \ j_{13} = \bfS' \, ; & & 
   {\rm (4a)} \ -j_{23}-1 = \bfS \comma & {\rm (4b)} \ -j_{23}-1 = \bfS'
     \, ;
   \end{array}
     \label{poleQ}
\eqe
where $\bfS' = -1-b^{-2}-\bfS$. 
The structure of the poles here can be different
from that in (\ref{j3pol}), since we are considering 
highest weight representations. Note that the $\hat{sl}(2)$ representation 
is largely reduced by highest weight conditions. 

\vskip 3ex
\noindent
{\bf 7.} \quad  
Now, we would like to consider
the OPE involving the degenerate highest weight representation 
whose spin is given by
\eqb
   2j +1 &=&  {\rm (I)} \ \    (l+1) + n b^{-2} \quad \mbox{or}
         \quad 
         {\rm (II)} \ \ -(l+1) - (n+1) b^{-2} \comma
      \label{degenerate}
\eqe
with $l,n \in \bfZ_{\geq0}$. 
From the representation theory of the current algebra, one can derive that 
the OPE can be non-vanishing among 
a degenerate primary with spin $j_1$ in (\ref{degenerate}),
some generic primary with spin $j_2$, and primaries in 
highest weight representations with the following spin $j_3$ 
\cite{Awata-Y}:
\eqb
\begin{array}{rrclll}
  {\rm (Ia)} & j_3 &=& j_2-j_1+u+wb^{-2} & (0\le u\le l,   & 0\le w\le n),\\
  {\rm (Ib)} & j_3 &=& j_1-j_2-u-wb^{-2} & (1\le u\le l+1, & 1\le w\le n),\\
  {\rm(IIa)} & j_3 &=& j_1-j_2+u+wb^{-2} & (0\le u\le l,   & 0\le w\le n),\\
  {\rm(IIb)} & j_3 &=& j_2-j_1-u-wb^{-2} & (1\le u\le l+1, & 1\le w\le n).
\end{array}
\label{degfu}
\eqe
Here, the first and second sequences correspond to case (I) 
in (\ref{degenerate}) and the third and fourth  to (II).
Note that these are not invariant under exchanging $j_3$ and $-j_3-1$
and they represent non-equivalent cases, since we are considering
the highest weight representations with $m_{\rm max} = j_3$. 

 This OPE would be analyzed also by using the formula (\ref{OPE}).
 When one of the two operators in the product has spin $j$
 as given in (\ref{degenerate}), $D(j_a)$ vanishes for generic
 values of $j_{2,3}$ due to a factor $\Up(-2j_1b)$ in the numerator.
However, by a careful analysis,  we find that  
there are still some contributions to
 the right hand side of the OPE from pairs of poles
 pinching the contour and degenerating into double poles.
In this process, we first need to separate such pairs of poles 
infinitesimally, and then take the pinching limit. There are other 
cases in which two poles are colliding and compensate  
the zero from $\Up(-2j_1b)$. However, in those cases, 
two contributions have opposite signs and cancel each other. 

 From the table of the poles in (\ref{j3pol}), 
 we find the contributions from the pinching poles at
\eqb
\begin{array}{crcllll}
 \mbox{(1a - 3a)}& 
  j_3 &=& j_2-j_1+u+wb^{-2} & (0\le u\le l,   & 0\le w\le n),\\
 \mbox{(2a - 4b)}& 
  j_3 &=& j_1-j_2-u-wb^{-2} & (1\le u\le l+1, & 0\le w\le n),\\
 \mbox{(2b - 4a)}& 
  j_3 &=& j_1-j_2+u+wb^{-2} & (0\le u\le l,   & 0\le w\le n+1),\\
 \mbox{(1b - 3b)}& 
  j_3 &=& j_2-j_1-u-wb^{-2} & (1\le u\le l+1, & 0\le w\le n+1).
\end{array}
 \label{degPhij}
\eqe
The first column stands for the sequences in (\ref{j3pol}) 
containing the pinching poles.
The first two cases correspond to case (I) and the last two to case (II).
This result is slightly different from (\ref{degfu}) 
(and correct the statement in \cite{Teschner1}). 
Tracing the disagreements, we find
 that the additional pinchings all originate from the ``contact terms''
 which appear in the right hand side of (\ref{OPE}) multiplied
 by derivatives of $\delta^2(x_1-x_2)$ via (\ref{res}).
If we work with $\Phi^j_{m\mbar}$, not $\Phi_j$, 
 this discrepancy is expected to be resolved, since we are considering 
highest weight representations. In fact, in the OPE for $\Phi^j_{m\mbar}$ 
in the previous section, such contact terms did not appear, though some 
of the poles associated with them remain in (\ref{poleQ}) in disguise.

Thus, let us consider the above OPE following the formulation 
for $\Phi^j_{m\mbar}$, namely, (\ref{PhijmmOPE}) and (\ref{Qja}). 
The procedure is the same.
The contributions to the OPE come only from the pinching poles. 
This assures that the $m$-dependent poles 
are irrelevant in our argument as long as $\Phi^{j_2}_{m_2\mbar_2}$ is
generic, although the prescription for the $m$-dependent poles is yet 
to be determined.

From (\ref{poleQ}),
we then find the table of the allowed $j_3$:
\eqb
   \begin{array}{crclll}
     \mbox{(1a - 3a)} & j_3 &=& j_2-j_1 + u + w b^{-2} 
            & (0 \leq u \leq l \comma & 0 \leq w \leq n) \comma \\ 
     \mbox{(2a - 4a)} & j_3 &=&  j_1 - j_2 - u - w b^{-2} 
            & (1 \leq u \leq l+1 \comma & 0 \leq w \leq n) \comma \\
     \mbox{(2b - 4b)} & j_3 &=& j_1 - j_2 + u + w b^{-2} 
            & (0 \leq u \leq l \comma & 1 \leq w \leq n) \comma \\ 
     \mbox{(1b - 3b)} & j_3 &=&  j_2 - j_1 - u - w b^{-2} 
            & (1 \leq u \leq l+1 \comma & 1 \leq w \leq n) \period 
   \end{array}
  \label{j3Q}
\eqe 
In our formulation, spin $j$ and $-j-1$ appear always in pairs, and they 
are regarded as giving equivalent representations with common 
$m$ and $\mbar$ as in (\ref{Phi-Phi}).
The first and second, and the third and fourth sequences represent
such pairs. To compare this table with the results in \cite{Awata-Y} in which 
the primaries with $j_3$ belong to highest weight representations, we 
further need to impose highest weight conditions, 
taking into account the difference of conventions.
Denoting a pair of $j_3$'s by $j_3^{(1)}$ and 
$j_3^{(2)} = - j_3^{(1)} -1$, the condition for $m_3$ reads, e.g., 
$ m_3 = j^{(1)}_3 - \bfZ_{\geq 0}$ (and similarly for $\mbar_3$). 
Since the primary with $j_2$ is in a generic representation and $ m_3$ 
is determined by the conservation of $ m $, i.e.,  
$ m_3 = m_1 + m_2 $, we can always choose 
$ m_2 $ so that the condition is satisfied. 
$ m_3$ is then fixed for given $\Phi^{j_1}_{m_1\mbar_1}$ and 
$\Phi^{j_2}_{m_2\mbar_2}$. In turn, this means that 
the contribution from $j_3^{(2)}$ vanishes, 
because $\Gamma^{-1}(j_3^{(2)}+ m_3+1) = 0$ in $Q$. 
We note that either $j$ or $-j-1$ was selected also in an observation 
in section 4 after imposing highest weight conditions.
    
From the two tables (\ref{degfu}) and (\ref{j3Q}), we find that 
(1a-3a) $\sim$ (2a-4a) agrees with (Ia), and (2b-4b) $\sim$ 
(1b-3b) with (IIb). The results in 
\cite{Awata-Y} give the possible representations which are allowed 
in the OPE, and all of them do not necessarily appear. For example, 
correct gluing of the left and right sector may restrict them. 
In fact, in the case with both
$j_1$ and $j_2$ in (\ref{degenerate}), there are examples in which 
only a part of the results in \cite{Awata-Y} appears 
in concrete models \cite{Petersen-RY,FGP}.\footnote{
In the subsequent paper, the authors of \cite{Petersen-RY} showed that 
the full results of \cite{Awata-Y} can be obtained by a careful analysis of
integration contours and screenings \cite{Petersen-RY2}. The full results
have also been obtained in \cite{Andreev,FGP}.   
Regarding the OPE for the degenerate representations,
see also \cite{degOPE}. We would like 
to thank J. Rasmussen, O. Andreev, P. Furlan and V. Petkova for 
bringing our attention to these references and useful comments.
} 
Our results are analogous, and they are regarded as consistent 
with \cite{Awata-Y}. This was not the case for the results 
in (\ref{degPhij}) which was obtained by the OPE of $\Phi_j$. 
In this way, a puzzle about the OPE using $\Phi_j$ is resolved.

\vskip 3ex
\noindent
{\bf 8.} \quad  
We have analyzed the properties of $\Phi^j_{m\mbar}$, 
and calculated their correlation functions.
Using these correlators, we have considered the OPE in the case including 
the degenerate highest weight representation, and obtained the correct 
OPE. 

Our arguments indicate an importance of $\Phi^j_{m\mbar}$ 
in considering highest weight representations, and 
further support the basic idea in \cite{Teschner1,Teschner2} 
that the OPE for the models with affine $SL(2)$ symmetry is obtained 
by continuations from the $H_3^+$ case. We also saw that 
$\Phi^j_{m\mbar}$ represent correct wave functions on $SL(2,R)$ 
after imposing highest weight conditions and continuing the parameters.
Thus, we expect that the OPE in the $SL(2,R)$ WZNW model, 
which is of considerable interest, is obtained along this approach.
Here, we encounter important problems yet to be clarified: One is how to 
deal with the $m$-dependent poles. The other is how to incorporate the
spectral flowed sectors which would play a significant role in the $SL(2,R)$
case \cite{Maldacena-O}. For the former problem, 
the representation theory of $SL(2,R)$ may be a good guide. In fact, one
can give a prescription, by hand,  so that the OPE becomes consistent 
with the representation theory.  
However, these problems need further substantial works, and  
they are beyond the scope of this short note.  

\vskip 6ex
\ni
{\it Note added} 

 While the first version of the manuscript was being written, 
a paper \cite{Giribet-N2} appeared which has some overlap with 
our discussions. After the first 
version was written, the OPE of the $SL(2,R)$ WZNW model has 
been discussed in \cite{YS,Maldacena-O2}.

\vskip 6ex
\ni
{\it Acknowledgments}

We would like to thank H. Awata, T. Eguchi, T. Fukuda, 
K. Hamada, N. Ishibashi, 
S. Mizoguchi, N. Ohta, K. Okuyama, J. Rasmussen, Y. Yamada and S.K. Yang
for useful discussions and correspondences.  
The work of K.H. was supported in part
 by JSPS Research Fellowships for Young Scientists, whereas
the work of Y.S. was supported in part  
by Grant-in-Aid for Scientific Research on Priority Area 707 and
Grant-in-Aid for Scientific Research (No.12740134) from
the Japan Ministry of Education, Culture, Sports, Science and Technology.

\newpage
\def\thebibliography#1{\list
 {[\arabic{enumi}]}{\settowidth\labelwidth{[#1]}\leftmargin\labelwidth
  \advance\leftmargin\labelsep
  \usecounter{enumi}}
  \def\newblock{\hskip .11em plus .33em minus .07em}
  \sloppy\clubpenalty4000\widowpenalty4000
  \sfcode`\.=1000\relax}
 \let\endthebibliography=\endlist
\vskip 5ex
\begin{center}
 {\bf References}
\end{center}
\vskip 1ex

\par


\begin{thebibliography}{999}
\parskip=-3pt
%
%
\bibitem{Gawedzki} K.~Gaw{\c e}dzki,
   Nucl.  Phys. B {\bf 328} (1989) 733; \
    NATO ASI: Cargese 1991: 0247-274.
%
\bibitem{Teschner1} J.~Teschner,
     Nucl.  Phys. B {\bf 546} (1999) 390; \ 
   Nucl. Phys. B {\bf 546} (1999) 369.
%
\bibitem{Teschner2} J.~Teschner, 
      Nucl. Phys. B {\bf 571} (2000) 555.
%
\bibitem{Giveon-K} 
   A.~Giveon and D.~Kutasov,
    JHEP {\bf 9910} (1999) 034; \
    JHEP {\bf 0001} (2000) 023.
%
\bibitem{Maldacena-O} J.~Maldacena and H.~Ooguri,
   J. Math. Phys.  {\bf 42} (2001) 2929; \
   J.~Maldacena, H.~Ooguri and J.~Son,
   J. Math. Phys.  {\bf 42} (2001) 2961.
%
\bibitem{deBoer-ORT} 
J.~de~Boer, H.~Ooguri, H.~Robins and J.~Tannenhauser,
 \jhep{9812}{1998}{026}.
%
\bibitem{Kutasov-S} D.~Kutasov and N.~Seiberg,
     JHEP {\bf 9904} (1999) 008.
%
\bibitem{Ishibashi-OS} N.~Ishibashi, K.~Okuyama and Y.~Satoh,
     Nucl. Phys. B {\bf 588} (2000) 149.
%
\bibitem{Hosomichi-OS} K.~Hosomichi, K.~Okuyama and Y.~Satoh, 
      Nucl. Phys. B {\bf 598} (2001) 451.
%
\bibitem{Dorn-O} H. Dorn and H.J. Otto,
   Nucl. Phys. B {\bf 429}, 375 (1994). 
%
\bibitem{Zamolodchikov-Z} A.B.~Zamolodchikov and Al.B.~Zamolodchikov,
    Nucl. Phys.B {\bf 477} (1996) 577.
%
\bibitem{Bateman} Bateman Manuscript Project, 
   ``{\it Higher transcendental functions}'', Vol.~1, 
     McGraw-Hill, New York (1953).
%
\bibitem{Liu} H.~Liu, Phys. Rev. D {\bf 60} (1999) 106005.
%
\bibitem{Moore-SS} G.~Moore, N.~Seiberg and M.~Staudacher,
    Nucl. Phys. B {\bf 362} (1991) 665.
%
\bibitem{Kato-S} A.~Kato and Y.~Satoh,
   Phys. Lett. B {\bf 486} (2000) 306.
%
\bibitem{Becker-B} 
   K.~Becker and M.~Becker,
    Nucl. Phys. B {\bf 418} (1994) 206.
%
\bibitem{Awata-Y} H.~Awata and Y.~Yamada,
    Mod. Phys. Lett. A {\bf 7} (1992) 1185.
%
%
\bibitem{Petersen-RY} J.L.~Petersen, J.~Rasmussen and M.~Yu,
       Nucl. Phys. B {\bf 457} (1995) 309.
%
\bibitem{FGP} P. Furlan, A.Ch. Ganchev, R. Paunov and V.B. Petkova, 
    Nucl. Phys. B {\bf 394} (1993) 665; ~
    P.~Furlan, A.Ch.~Ganchev and V.B.~Petkova, 
   Nucl. Phys. B {\bf 491} (1997) 635.
%
\bibitem{Petersen-RY2} J.L.~Petersen, J.~Rasmussen and M.~Yu,
   Nucl. Phys. B {\bf 481} (1996) 577.
%
\bibitem{Andreev}  
       O.~Andreev, Phys. Lett. B {\bf 363} (1995) 166. 
%
\bibitem{degOPE} B. Feigin and F. Malikov,
        Lett. Math. Phys. {\bf 31} (1994) 315; \\ 
      A.Ch.~Ganchev, V.B.~Petkova and G.M.T.~Watts,
       Nucl. Phys. B {\bf 571} (2000) 457.
%
\bibitem{Giribet-N2} G.~Giribet and C.~N{\'u}{\~n}ez,
    JHEP {\bf 0106} (2001) 010.
%
\bibitem{YS} Y.~Satoh, 
   ``{\it Three-point functions and operator product expansion in the $SL(2)$ 
   conformal field theory}'', to appear in Nucl. Phys. B, \hpt{0109059}.
%
\bibitem{Maldacena-O2}
   J.~Maldacena and H.~Ooguri, 
   ``{\it Strings in $AdS_3$ and the $SL(2,R)$ WZW model. III: 
    Correlation  functions}'', \hpt{0111180}.
%
\end{thebibliography}
\end{document}